\documentclass[twocolumn,showpacs,aps,prl]{revtex4}
\usepackage{epsfig}

\begin{document}

\title{Measurement of the Temperature Dependence of the Casimir-Polder Force}
\author{J.~M. Obrecht$^{1,\star}$, R.~J. Wild$^{1}$, M. Antezza$^{2}$, L.~P. Pitaevskii$^{2,3}$, S. Stringari$^{2}$, E.~A. Cornell$^{1,\dag}$}
\affiliation{ $^1$JILA, National Institute of Standards and Technology and University of Colorado, Boulder, Colorado 80309-0440, USA
\\and Department of Physics, University of Colorado, Boulder, Colorado 80309-0390, USA
\\$^2$Dipartimento di Fisica, Universit\`{a} di Trento and CNR-INFM BEC Center, Via Sommarive 14, I-38050 Povo, Trento, Italy
\\$^3$Kapitza Institute for Physical Problems, ulitza Kosygina 2, 119334 Moscow, Russia}

\date{\today}

\begin{abstract}

We report on the first measurement of a temperature dependence of the Casimir-Polder force.  This measurement was obtained by positioning a
nearly pure $^{87}$Rb Bose-Einstein condensate a few microns from a dielectric substrate and exciting its dipole oscillation. Changes in the
collective oscillation frequency of the magnetically trapped atoms result from spatial variations in the surface-atom force.  In our experiment,
the dielectric substrate is heated up to 605~K, while the surrounding environment is kept near room temperature (310~K). The effect of the
Casimir-Polder force is measured to be nearly 3 times larger for a 605~K substrate than for a room-temperature substrate, showing a clear
temperature dependence in agreement with theory.

\end{abstract}


\maketitle

The Casimir force and its molecular cousin, the van der Waals force, are not only fascinating scientifically but also important technologically,
for example in atomic force microscopy and microelectromechanical systems.  Like the tension in a rubber band, the Casimir force is a
conservative force arising from microscopic fluctuations.  The Casimir force is also the dominant background effect confounding
attempts~\cite{ederth,kapitulnik,lamoreaux1997,decca2005} to set improved limits on exotic forces at the $10^{-8}$~m to $10^{-5}$~m length
scale; progress towards a deeper understanding is valuable in that context.   Typically one uses ``Casimir"~\cite{casimir1948} to refer to the
force between two bulk objects, such as metallic spheres or dielectric plates, and ``Casimir-Polder" (CP)~\cite{cp1948} to describe the force
between a bulk object and a gas-phase atom. The underlying physics~\cite{mauronew} is largely the same, however, and, particularly in the limit
of separations exceeding one micron, it can be more convenient to study the latter system due to ease in rejecting systematic errors such as
electrostatic patch potentials~\cite{jeffPRA, davePRA}.

The Casimir force arises from fluctuations of the electromagnetic field and is usually thought of as being purely quantum-mechanical. However, at
nonzero temperatures, the fluctuations also have a thermal contribution, which was investigated by Lifshitz~\cite{lifshitz}. Precise theoretical
modelling of Casimir forces takes into account effects such as surface roughness, finite conductivity, substrate geometry, and nonzero
temperature, but the latter term has never been unambiguously observed experimentally (see~\cite{CasCorr} and references therein). In earlier
Casimir~\cite{overbeek1978, lamoreaux1997, mohideen1998, capasso2001, ruoso2002, decca2003} and Casimir-Polder~\cite{hinds1993, aspect1996,
shimizu2001, dekieviet2003, vuletic2004, ketterle2004, shimizu2005, davePRA} experiments, thermal effects were predicted to be on the order of
experimental uncertainties or less because (a) the temperature of the apparatus could not be varied over a large range and (b) the experiments
worked over small separations compared to the wavelength of thermal radiation, where thermal corrections are small.

In this Letter, we report the first measurement of a temperature dependence of the Casimir-Polder force, indeed the first conclusive temperature
dependence of any Casimir-like system. A key feature of this work is that the apparatus temperature is spatially nonuniform.  This allows for an
experimental confirmation of an appealing theoretical insight: The thermal electromagnetic-field fluctuations that drive the CP force can be
separated into two categories --- those that undergo internal and those that undergo external reflection at the surface. These two categories of
fluctuations contribute to the total force with opposite sign; in thermal equilibrium, they very nearly cancel, masking the underlying scale of
thermal effects. Working outside thermal equilibrium, we observe thermal contributions to the CP force that are 3 times as large as the
zero-temperature force.

To review the main regimes in surface-atom forces:  For a surface-atom separation $x$ much less than the wavelength of the dominant resonances
in the atom and substrate, the potential $U$ scales as $1/x^3$ (van der Waals-London regime).  At longer distances, retardation effects cause a
crossover to $U \sim 1/x^4$ (Casimir-Polder). At still longer distances when $x$ is comparable to the blackbody peak at temperature $T$,
temperature effects become important, and in thermal equilibrium ($T=T_S=T_E$, as defined below), there is a second crossover, back to $U \sim
T/x^3$ (Lifshitz).

\begin{figure}
\leavevmode \epsfxsize=3.0in \epsffile{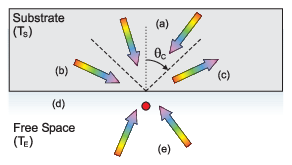} \caption{\label{fig:radiation} (Color online) Cartoon drawing of thermal fluctuations
near the surface of a dielectric substrate (shaded region). (a) Internal radiation striking the surface at angles less than the critical angle
$\theta_{C}$ does not contribute to the Casimir-Polder force. However, internal radiation impingent at larger angles (b) undergoes total
internal reflection (c) and contributes to an overall AC Stark shift by creating evanescent waves in free space (d). Surrounding the atom (red
circle) is radiation from the environment (e) which contributes to the CP force by creating standing waves at the surface.  The force does not
arise from radiation pressure but rather from gradients in intensity.  The surface-atom force becomes more attractive for $T_S > T_E$ and more
repulsive for $T_S < T_E$.}
\end{figure}

Recent theoretical work~\cite{mauroNEQ} has shown that thermal corrections to the Casimir-Polder force are separable into those arising from
thermal fluctuations within the substrate at temperature $T_S$~\cite{henkel} and those arising from radiation impinging from presumed distant
walls at an environmental temperature $T_E$. At $T_S\neq$0, electromagnetic fluctuations from within the surface have an evanescent component
that extends into the vacuum with maximum intensity at the surface, giving rise to an attractive AC Stark potential (Fig.~\ref{fig:radiation}).
External radiation at $T_E$, impinging at different angles, reflects from the substrate surface, giving rise to a field distribution whose
intensity falls smoothly to a minimum at the substrate surface. The resulting Stark shift from the external radiation then pulls the atom
$\it{away}$ from the surface, contributing a repulsive term to the potential~\cite{absorption}. Antezza $\textit{et al}$.~\cite{mauroNEQ}
recently predicted that the nonequilibrium contribution to the CP potential asymptotically scales as $U_{NEQ}\sim(T_{S}^2-T_{E}^2)/x^2$.  This
novel scaling dependence dominates at long range. One can thus temperature-tune the magnitude of this long-range force and, in principle, even
change the sign of the overall force.

We observe the temperature dependence of the Casimir-Polder force between a rubidium atom and a dielectric substrate by measuring the collective
oscillation frequency of the mechanical dipole mode of a Bose-Einstein condensate (BEC) near enough to a dielectric substrate for the CP force
to measurably distort the trapping potential.  This distortion of the trap results in changes to the oscillation frequency proportional to the
gradient of the force:

\begin{equation}
\gamma_{x} \equiv \frac{\omega_{o}-\omega_{x}}{\omega_{o}} \simeq \frac{1}{2m\omega_{o}^{2}}~\langle\partial_{x}F_{CP}\rangle,
\end{equation}

\noindent where $m$ is the mass of the $^{87}$Rb atom, and $\gamma_{x}$ is defined as the fractional frequency difference between the
unperturbed trap frequency $\omega_{o}$ and $\omega_{x}$, the trap frequency perturbed by the CP force $F_{CP}$.

The use of a BEC in this work is not conceptually central.  The force between the substrate and the condensate is the simple sum of the force on
the individual atoms of the condensate.  For our purpose, the condensate represents a spatially compact collection of a relatively large number
of atoms whose well-characterized Thomas-Fermi density profile facilitates the spatial averaging and the inclusion of nonlinear effects in the
oscillations, necessary for the quantitative comparison between theory and experiment~\cite{mauroEQ,davePRA}.

Experimental details, surface-atom measurement and calibration techniques, along with a detailed discussion of measurements of stray electric
and magnetic fields appear in~\cite{daveJLTP, jeffPRA,davePRA}. In brief, the experiment consists of $2.5\times 10^5$ $^{87}$Rb atoms
Bose-condensed (condensate purity $>0.8$) in the $|F=1,m_{F}=-1\rangle$ ground state.  The condensate is produced $\sim$1.2 mm below a
dielectric substrate in a Ioffe-Pritchard-style magnetic trap (trap frequencies of 229 Hz and 6.4 Hz in the radial and axial directions,
respectively), resulting in respective Thomas-Fermi radii of 2.69$~\mu$m and 97.1$~\mu$m.

\begin{figure}
\leavevmode \epsfxsize=3.0in \epsffile{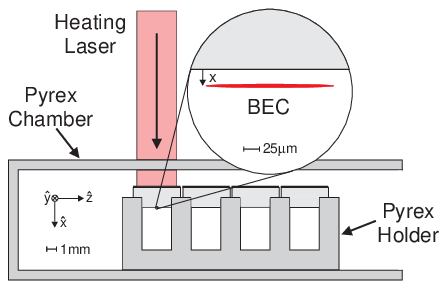} \caption{\label{fig:apparatus} (Color online) Side view of the apparatus.  Shown is a
scale drawing of the fused silica substrate (left-most of the four substrates) with a top layer of graphite. The graphite absorbs the light from
the laser, heating the substrate. The pyrex holder is isolated enough from the substrate to allow a hot substrate -- cool environment scenario.
The enlargement in the inset shows the BEC at a distance $x$ from the surface.}
\end{figure}

The dielectric substrate studied consists of uv-grade fused silica $\sim2\times8\times5$ mm$^3$ in size ($x,y,z$ directions, respectively)
sitting atop a monolithic pyrex glass holder inside a pyrex glass cell which composes the vacuum chamber (Fig.~\ref{fig:apparatus}). The top
surface ($-\hat{x}$~face) of the substrate is painted with a $\sim$100 $\mu$m thick opaque layer of graphite and treated in a high-temperature
oven prior to placement in the vacuum chamber.  The observed lifetime of the BEC places a strong, robust, upper bound on the total pressure of
residual gas just below the substrate surface of $\sim3\times 10^{-11}$ torr, even at $T_{S}$ = 605~K.

The fused silica substrate was heated by shining $\sim$1 W of laser light (860~nm) on the graphite layer.  The rough texture of the pyrex holder
creates near point contacts with the substrate corners, providing good thermal isolation between the holder and the substrate. This technique
allows us to vary the temperature of the substrate while maintaining near room temperature vacuum chamber walls and only slightly elevated
holder temperatures.

\begin{figure}
\leavevmode \epsfxsize=3.0in \epsffile{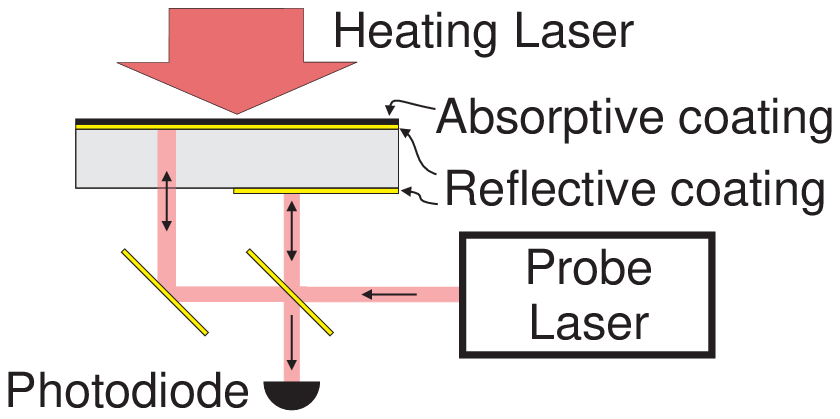} \caption{\label{fig:interferometer} (Color online) Interferometric temperature
measurement apparatus (sizes not to scale). Shown is a schematic view of the Michelson interferometer used to non-perturbatively measure the
substrate temperature.  As the substrate heats, the glass both expands $\textit{and}$ changes its index of refraction, creating an
interferometric signal.}
\end{figure}

\begin{figure}
\leavevmode \epsfxsize=3.0in \epsffile{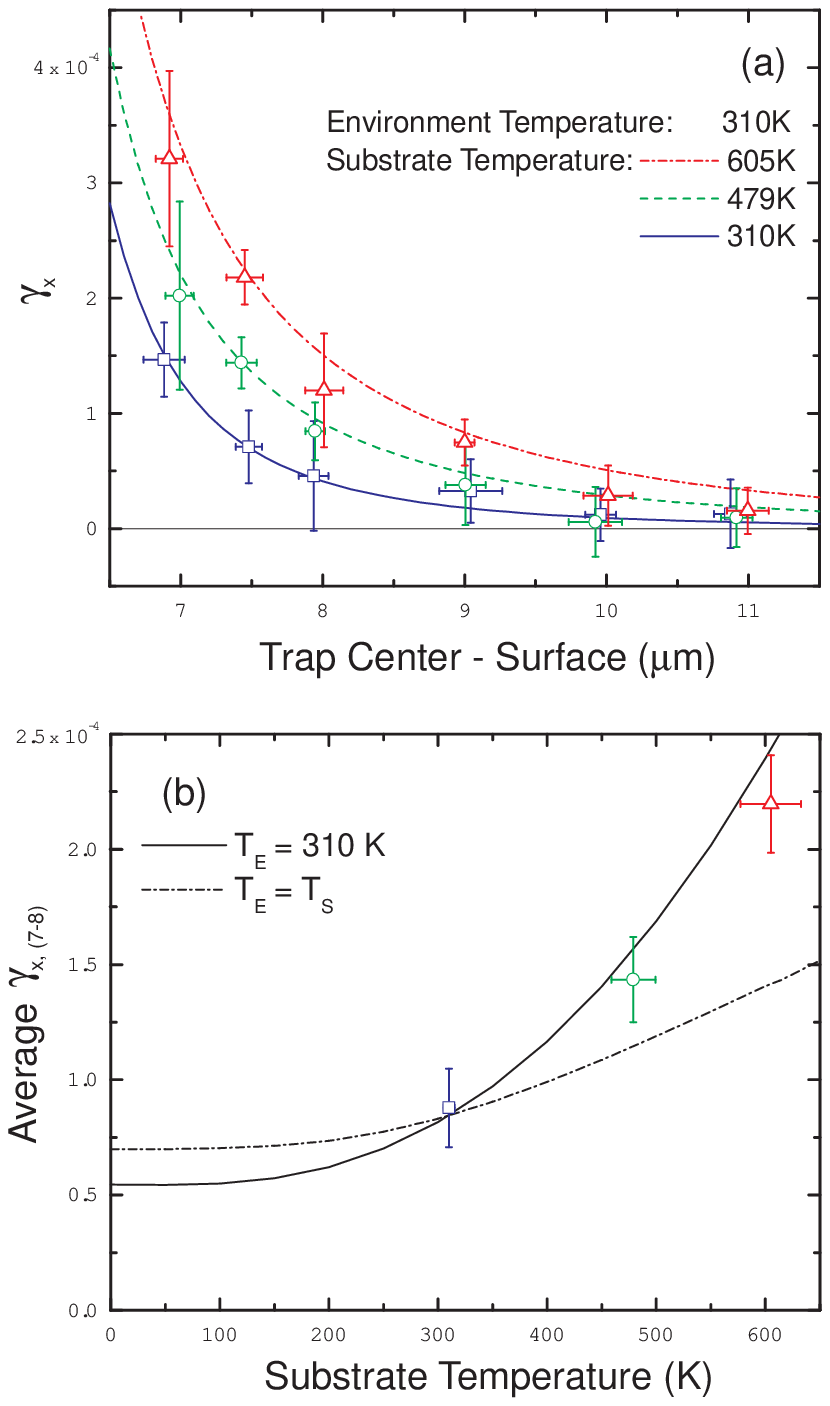} \caption{\label{fig:data} (Color online) (a) Fractional change in the trap frequency
due to the Casimir-Polder force. Pictured are three sets of data and accompanying theoretical curves with no adjustable parameters for various
substrate temperatures. The blue squares represent data taken with a 310 K substrate; green circles, a 479 K substrate; and red triangles, a 605
K substrate.  The environment temperature is maintained at 310 K. The error bars represent the total uncertainty (statistical and systematic) of
the measurement. (b) Average values of $\gamma_{x}$ from (a) (for trap center to surface positions 7.0, 7.5, and 8.0 $\mu$m) plotted versus
substrate temperature, demonstrating a clear increase in strength of the CP force for elevated temperatures.  The solid theory curve represents
the nonequilibrium effect (corresponding 7--8 $\mu$m average), while the dash-dot theory curve represents the case of equal temperatures.}
\end{figure}

The temperature of the fused-silica substrate as a function of the heating laser power was determined in an offline calibration apparatus,
constructed to be a near-identical version of the main vacuum chamber, except with improved optical access for a temperature probe laser. The
probe laser is coherently split between two arms of a Michelson interferometer (Fig.~\ref{fig:interferometer}).  The resulting fringe shifts are
proportional to changes in substrate temperature. A finite-element numerical model of the thermal system agreed with our measurements and
contributed to our confidence that the temperature of the substrate and the environment were understood. Residual systematic uncertainties in
temperature are reflected in the error bars in Fig.~\ref{fig:data}(b).

The experiment, described in detail in~\cite{davePRA}, begins with an adiabatic displacement of the atom cloud to a distance $x$ from the bottom
surface ($+\hat{x}$ face) of the substrate via the addition of a vertical bias magnetic field.  The cloud is then resonantly driven into a
mechanical dipole oscillation by an oscillatory magnetic field~\cite{amplitude}.  After a period of free oscillation the relative position of
the cloud is determined by destructive imaging after $\sim$5 ms of anti-trapped expansion. This process is repeated for various times in the
free oscillation.  The center-of-mass position is recorded at short times to determine the initial phase of the oscillation and at later times
($\sim$1s) to precisely determine the oscillation frequency.  Data is taken consecutively alternating the trap center position between a
distance $x$, close to the surface, and a normalizing distance $x_{o}=15~\mu$m.  The distance $x_{o}$ is sufficiently far from the substrate to
avoid surface perturbations, yet close enough to provide a local oscillator $\omega_{o}$ with which the data taken at $x$ can be compared. Data
sets are then taken at a number of surface-atom positions (between 7--11~$\mu$m) and for various substrate temperatures (310, 479 and 605 K,
taken in random order, several times, and averaged in Fig.~\ref{fig:data}).

The results in Fig.~\ref{fig:data}(a) show the fractional change in the trap frequency $\gamma_{x}$ plotted as a function of the trap center to
surface position $x$. The blue squares show the measured effect of the room-temperature Casimir-Polder force [$T_{S}$=310(5)~K] on the trap
frequency.  The increase in the strength of the CP force due to thermal corrections becomes obvious when the substrate is heated to $479(20)$~K
(green circles) and even more pronounced when it is at $605(28)$~K (red triangles).  These measurements were all done maintaining a room
temperature environment for which the pyrex vacuum chamber walls were measured to be $T_{E}=310(5)$~K.  The curves in Fig.~\ref{fig:data}(a)
represent the theoretical predictions~\cite{mauroNEQ} for corresponding substrate-environment temperature scenarios, showing excellent agreement
with the measurements.

For statistical clarity in data analysis, the average value of $\gamma_{x}$ can be computed for each substrate temperature (using trap center to
surface positions 7.0, 7.5, and 8.0 $\mu$m only).  These values, plotted in Fig.~\ref{fig:data}(b), clearly show a significant increase in the
strength of the Casimir-Polder force for hotter substrate temperatures; they also distinguish the nonequilibrium theory (solid) curve from the
equilibrium (dash-dotted) curve, for which a much smaller force increase is predicted.

The killer systematic in Casimir force experiments is often stray electric fields caused by poorly characterized surface properties.  We put
great care into $\textit{in situ}$ characterizing, for the 605~K and 310~K temperature scenarios in Fig.~\ref{fig:data}, stray magnetic fields
and gradients~\cite{gradients} of stray electric fields, using techniques we developed in Ref.~\cite{jeffPRA,davePRA,elecpaper}. From the
magnitude of near-surface dc electric fields, we estimate the surface density of adsorbed alkali atoms to be much less than 1/1000 of a
monolayer at all measured temperatures, far too low to change the optical properties of the substrate by the factors of nearly 3 that we see the
Casimir-Polder force change by. In addition, the substrate is optically flat at visible wavelengths. Therefore, at the much longer relevant
length scales of our experiment any residual surface roughness will be negligible. We can also rule out that the measured increase in strength
of the CP force comes from a change in the dielectric constant with increasing temperature, or mechanical effects on the atoms from the heating
laser.

One also must consider the quality of the blackbody radiation emitted by the environment. While the pyrex walls of the chamber are transparent
at visible and near-infrared wavelengths, at 5--10 $\mu$m wavelengths the walls are opaque, with an emissivity $>$ 0.8.

In conclusion, we have made the first measurement discerning the temperature dependence of the Casimir-Polder force and confirmed the
nonequilibrium theory of Antezza $\textit{et al}$.~\cite{mauroNEQ}. The strength of this force was shown to increase by a factor of nearly 3 as
the substrate temperature doubles.

We are grateful for experimental assistance from C.~Gillespie and useful discussions with C.~Henkel, G.~Roati, D.~Harber, and the JILA ultracold
atoms and molecules collaboration.  This work was supported by grants from the NSF and NIST and is based upon work supported under a NSF Graduate
Research Fellowship.


\begin{thebibliography}{10}
\bibitem{absorption}    
Here we assume that the atom does not absorb radiation ($T_{E}$ and $T_{S}$ are small with respect to atomic resonances).
\bibitem{amplitude}     
The amplitude of the oscillation is carefully calibrated to be $2.50(2)~\mu$m in the $\hat{x}$ direction for all measurements.
\end{thebibliography}
\end{document}